\documentclass[preprint,showpacs,preprintnumbers,amsmath,amssymb]{revtex4}

\usepackage{graphicx}% Include figure files
\usepackage{dcolumn}% Align table columns on decimal point
\usepackage{bm}% bold math
\usepackage{amsmath}
\usepackage{amssymb}
\usepackage{latexsym}
\usepackage{epsfig}
\usepackage{amsbsy}
\usepackage{array}
\usepackage{amssymb}
\usepackage{setspace}
\usepackage{bm}

\usepackage{CJKutf8}

\def\sint{\ifmmode{- \!\!\!\!\!\! \int}
	\else{\hbox{$- \!\!\!\! \int \ $}}\fi}

\begin{document}
\begin{CJK}{UTF8}{gbsn}

	%\preprint{Physical Review Letters}
	
	%\title{Non-conjugate Ghost Imaging and Its Ability Against Turbulence }% Force line breaks with \\
	\title{{ Ghost Aperture Synthesis Imaging with Computational Aberration Cancellation}}% Force line breaks with \\
	\author{Shuai Sun$^{1,2,3}$, Zhen-Wu Nie$^{1,2}$, Yue-Gang Li$^{2,4}$, Hui-Zu Lin$^{1,2,3}$, Wei-Tao Liu$^{1,2,3,*}$,Ping-Xing Chen$^{1,2,3}$}
	
	%email[wtliu@nudt.edu.cn]{Your e-mail address}
	%\thanks{}
	\affiliation{1,Institute for Quantum Science and Technology, College of Science, National University of Defense Technology, Changsha, Hunan,  410073, People's Republic of China}
	\affiliation{2,Interdisciplinary Center of Quantum Information, College of Science, National University of Defense Technology, Changsha, Hunan,  410073, People's Republic of China}
	\affiliation{3,Hunan Key Laboratory of Mechanism and Technology of Quantum Information, Changsha, Hunan,  410073, People's Republic of China}
	\affiliation{4,State Key Laboratory of Advanced Optical Communication Systems and Networks, Institute for Quantum Sensing and Information Processing, School of Sensing Science and Engineering, Shanghai Jiao Tong University, Shanghai, 200240, People's Republic of China}
	\affiliation{*wtliu@nudt.edu.cn}
	\date{\today}
	%\author{ \footnote{Author to whom correspondence should be
	%\href{mailto:wtliu@nudt.edu.cn} 
	%\affiliation{Lisi University, USA}
	
	\begin{abstract}
		
		Although optical aperture synthesis has been generally regarded as the only access to very large imager for over a century, the problem of phasing all the giant sub-apertures on the scale of wavelength is still prohibitive. Besides, the accompanied adaptive optics combatting the atmospheric turbulence is also bulky and complicated. We here propose a new paradigm aperture synthesis imager through turbulence, based on computational ghost imaging method. The complex aberrations on the signal path are computationally cancelled by introducing an optimum compensation phase on the reference path. With the advanced aberration cancellation, our imager is free from phasing and aberrations problem. The image degradation due to turbulence is also suppressed and even eliminated without any guide star or wavefront shaping device. Experimentally, diffraction-limited imaging is achieved under turbulence featuring transverse coherence length far less than the optical aperture of the system.
		
	\end{abstract}
	
	\maketitle
	
	Around 1890, H. Fizeau and A. A. Michelson successively came up with stellar interferometry\cite{Johnson1974,Labeyrie2006,Saha2002}, hence reveal that widely separated but coherently combined sub-apertures can realize the resolving ability far surpassing individual one. Since manufacturing a whole aperture with diameter of meters is much beyond the reach of current technologies, the technique of optical aperture synthesis (OAS) is recognized as the only approach to huge interferometers and imagers. The concept of OAS subsequently encouraged the birth of intensity interferometer\cite{HBT1956,Boal1990,Yabashi2001}, dominated the design of the hypertelescopes\cite{Labeyrie1999}, and inspired the blossoming field of radio astronomy\cite{Moreira2013}. 
	
	To achieve the diffraction limit of OAS systems, all the giant, separated sub-apertures need to be phased to an accuracy of optical wavelength, which requires various of specific alignment techniques including extremely accurate clocks and time markers\cite{Johnson1974,Labeyrie2006,Saha2002,Giovannetti2001}. In addition, bulky adaptive optics is also indispensable to actively eliminate the aberrations of OAS system caused by the gravity and temperature variation\cite{Labeyrie2006,Roggemann1997}, also to correct the wavefront distortion induced by atmospheric turbulence\cite{Roggemann1997, Gruneisen2021}. Either phasing problem or adaptive optics is a formidable task currently. Therefore, the OAS technique is usually of astronomical cost. Essentially, both of the phasing operation and adaptive optics are to preserve the first-order coherence of the photons from the sub-apertures. Utilizing the second-order coherence\cite{Twiss1957}, intensity interferometer can work without phasing operation. However, its sampling result only represents the squared magnitude of the Fourier transform of the image intensity. The loss of phase information severely obstructs the restoration of images, especially for complicated or faint objects\cite{Zernike1938,Liu2021}.
	
	%Simplifying to cancel the aberration among the sub-apertures and that caused by turbulence will make OAS more accessible. However, the expectation seems extravagant under the confocal plane imaging (CPI) scheme, since the first-order coherence of the light field guaranteed by both of the cophase accuracy and the adaptive optics process is a prerequisite to achieve the diffraction limit of CPI.  refers to    Around 1890, H. Fizeau and A. A. Michelson successively came up with the idea of stellar interferometry
	%

	%To consisted of a glass plate in one beam and, in the other, a pair of glass wedges which could slide laterally on one another to form a parallelsided plate with variable thickness}  the number of subapertures is still quite modest. 
	Ghost Imaging(GI) is an unconventional method to acquire images from the second-order coherence of light field\cite{Zhang2005,Valencia2006,Erkmen2010,Shapiro2012,Yu2016,Lis2018,Hodgman2019}. As a direct descendant, computational GI (CGI)\cite{Shapiro2008} also offers the advantage to perform standoff sensing lenslessly\cite{Hardy2013}, makes neither the first-order coherence of light field nor huge telescope necessary for diffraction-limited imaging\cite{Chen2009}. The wavefront distortion caused by atmospheric turbulence will apparently degenerate the performance of GI\cite{Hardy2013,Cheng2009,Dixon2011,Chan2011,Hardy2011}. This is accompanied with a good news that nonlocal wavefront tailoring is demonstrated recently. Exploiting the correlations displayed by the two-photon state, the dispersion\cite{Franson1992,Shapiro2010,Sensarn2010,Torres2012,Nodurft2020},  distortion\cite{Defienne2019,Lib2020} and aberration\cite{Black2020,Bhusal2021,Valencial2021} imposed on the local wave packet of one photon can be corrected by modulating the other counterpart photon. This therefore allows to perform (the analog of) two-photon aberration cancellation scheme based on GI system, and thereby construct a profoundly different OAS imager.
	
	Here we propose a new physical paradigm of OAS imager based on CGI\cite{Shapiro2008,Altmann2018} and nonlocal aberration cancellation. The unknown, intricate aberrations of the dephased sub-apertures and the wavefront deformation induced by turbulence on the signal path are both computationally suppressed and even cancelled, by introducing an optimum compensation phase on the reference path. This ghost OAS(GOAS) offers significant advantages, which are not only free from the phasing problem, but also capable to image through turbulence, without any guide star or wavefront sensing/shaping devices. Compared with conventional OAS, the phasing-free, sensor-less GOAS imager is far easier to construct and to conduct. The way of active illumination is also promising in the applications of standoff sensing.
	
	%in which an incoherent array of distributed sub-apertures emit phase-modulated laser to illuminate the object. In the presence of turbulence, image with the diffraction limit is computationally recovered after projecting sequential random speckle patterns and recording the resultant signals via a bucket detector.
	
	The GOAS imager is shown as Fig.1. A pulsed laser is divided into dozens of distributed, far separated sub-sources with parallel emitting direction. Each sub-source gets a relative phase due to the different beam length, which is random and unknown without any phasing operation. The phase of each sub-source is also modulated individually, thus temporally varying speckle patterns can be produced in the far field, which propagate through the atmospheric turbulence to a rough-surfaced target. The light reflected from the target is collected by a bucket detector. The sampling process of GOAS means projecting a sequence of random speckle patterns and recording the resultant bucket signals. When phase modulators and bucket detector with high bandwidth are used, the sampling can be finished within the coherence time of turbulence, which is typically milliseconds. Then, the relative phase can be treated as fixed within the sampling, under uncomplicated stabilization procedures for the whole source. In the presence of random relative phase and the turbulence, diffraction-limited images of the target can be recovered from the cross correlation between the bucket signals and the algorithmically optimized reference speckle patterns. 
	%The diffraction limit of the GOAS is determined by the baseline length of the synthetic source, which is far longer than the beam waist of each sub-source.
	
	%The GOAS imager is shown as Fig.1. A pulsed laser is divided into dozens of distributed s with parallel traveling direction. The phase of each is modulated randomly. Due to the different beam length, each transmitted beam gets a delayed phase, which is also random without any phasing operation. The beams emitted from the synthetic sources form speckles, propagate through the atmospheric turbulence to an extended, rough-surfaced target. The light reflected from the target is collected by a bucket detector. 
	%The diffraction-limited image can be reconstructed via an iterative algorithm performing on the results of the cross-correlation between the computed reference intensity pattern and the bucket signal.
	\begin{figure}[h]
		\centering
		\includegraphics[width=\columnwidth]{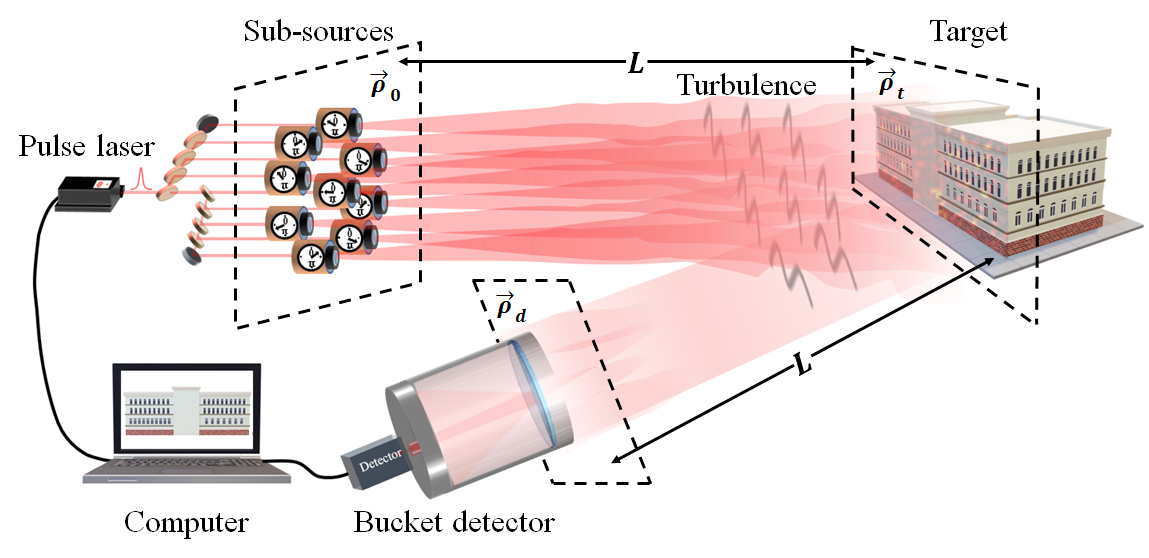}
		\caption{Diagrammatic sketch of GOAS. A laser is emitted from dozens of distributed, phase-modulated but dephased sub-sources. Temporally varying speckles undergoes the turbulence and illuminates the target. The reflected light is implemented bucket detection. Due to the relative phase among the sub-sources and the turbulence, the illumination patterns imprinted on the target are unknown. However, images with diffraction limit can be recovered after computational aberration cancellation.}
		\label{fig2} 
	\end{figure} 
	
	Taking the length difference among the divided beams less than the coherence length of the pulse, the $m^{th}$ \{$m=1,2,...M$\} in all $M$ illumination pattern emitted from the synthetic source can be written as 
	\begin{equation}
		E_0(\vec \rho_0, m)=\sum_{n=1}^{N}\xi(\vec \rho_0-\vec \rho_n)e^{i\left[\phi(n,m)+\phi_d(n)\right]}.
		\label{eq:1}
	\end{equation} 
	Here $N$ is the number of sub-sources.
	%the duration and repetition interval of which could be nanoseconds and the corresponding coherence length is tens of centimeters long. 
	$\vec\rho_0$ is the coordinates of the sources plane and $\vec \rho_n$ is the coordinates of the center of $n^{th}$ \{$n=1,2...,N$\} sub-source. $\phi(n,m)$ and $\phi_d(n)$ represents the modulation phase and relative phase, respectively.  The modulation phase temporally varies with the index $m$ while the relative phase keeps temporal constant during the sampling process. Besides, the modulation phase is known while the random relative phase is unknown. The Gaussian beam collimated by each sub-source is
	\begin{equation}
		\xi(\vec \rho_0-\vec \rho_n)=\sqrt{\frac{2P}{\pi \omega_0^2}}e^{-\frac{\left|\vec \rho_0-\vec \rho_n\right |^2}{2\omega_0^2}},
		\label{eq:2}
	\end{equation}
	where $P$ represents the laser intensity from each sub-source with units $Watts$. $\omega_0$ is the beam waist of the sub-source, which is several millimeters and far less than the coherence length of the turbulence. Since GOAS is opareated within a single atmospheric coherence time, the behavior of the light field $L$-m propagating through atmospheric turbulence can be characterized via extended Huygens-Fresnel principle\cite{Hardy2011,Hardy2013,Shapiro1974}, and the $m^{th}$ illumination pattern imprinted on the target and reflected back to the detector can be expressed by,
	\begin{equation}
		\begin{aligned}
			E_t(\vec \rho_t,m)& =\frac{k_0}{i2\pi L}\int_{a_0} d \vec \rho_0 E_0(\vec \rho_0,m) \\
			& \times e^{ik_0(L+\left|\vec \rho_t-\vec \rho_0\right|^2/2L)}e^{\Psi_t(\vec \rho_t,\vec \rho_0)}
		\end{aligned}.
		\label{eq:3}
	\end{equation}  
	and
	\begin{equation}
		\begin{aligned}
			E_d(\vec \rho_d,m)& =\frac{k_0}{i2\pi L} \int d \vec \rho_t E_t(\vec \rho_t,m)T(\vec \rho_t)\\
			& \times e^{ik_0(L+\left|\vec \rho_d-\vec \rho_t\right |^2/2L)}e^{\Psi_s(\vec \rho_d,\vec \rho_t)}.
		\end{aligned}.
		\label{eq:4}
	\end{equation} 
	respectively, where $a_0$ is the baseline length of the synthetic source and $a_0\gg \omega_0$. $k_0$ is the wave number. The target is a quasi-Lambertian reflector\cite{Hardy2011,Hardy2013} and its field-reflection coefficient is $T(\vec \rho_t)$. The complex function $\Psi_t(\vec \rho_t,\vec \rho_0)$ and $\Psi_s(\vec \rho_d,\vec \rho_t)$ specifies a frozen Kolmogorov-spectrum turbulence on the source-to-target and target-to-detector path, respectively. For instance, the real and imaginary parts of $\Psi_t(\vec \rho_t,\vec \rho_0)$ represent the log amplitude and phase fluctuations of the field from a point-like sub-source at $\vec \rho_0$, undergoing turbulence to receiver at $\vec \rho_t$.
	
	The arriving field from the target is collected by bucket detector and the produced current signal is 
	\begin{equation}
		i_b(m)= \frac{q\eta}{hf_0}\int d\vec \rho_d \mathcal{A}_b^2(\vec \rho_d)\left|E_d(\vec \rho_d,m) \right|^2
		\label{eq:7}
	\end{equation}
	in which $h$ is Planck’s constant and $f_0$ is the laser’s optical frequency in order that $i_b(m)$ have the correct units for photocurrent. $q$ is the charge of electron. $\eta$ and $\mathcal{A}_b(\vec \rho_d)$ represents the quantum efficiency and field-transmission pupil of the bucket detector, respectively. 
	
	Then an initial result of GOAS can be obtained from the cross-correlation between the bucket signal and reference intensity patterns as
	\begin{equation}
		G(\vec \rho_r)=\left < i_b(t)I(\vec \rho_r,t)\right >-\left < i_b(t)\right >\left < I(\vec \rho_r,t)\right >,
		\label{eq:10}
	\end{equation}
	with $\left < \cdot \right >$ being the temporal average of all $M$ samplings within a single atmospheric coherence time. $I(\vec \rho_r,m)$ is the $m^{th}$ reference intensity pattern (not shown in Fig.1), which is computed via a suppositional vacuum propagation of the synthetic source yields
	\begin{equation}
		I(\vec \rho_r,m)=\left |\int_{a_0} d \vec \rho_0^{~\prime} E_0^{~\prime}(\vec \rho_0^{~\prime},m)\frac{k_0}{i2\pi L}e^{ik_0(L+\left|\vec \rho_r-\vec \rho_0^{~\prime}\right |^2/2L)}\right |^2.	\label{eq:8}
	\end{equation}
	If all the sub-sources are phased, which means $E_0^{~\prime}(\vec\rho_0,m)=E_0(\vec\rho_0,m)$, then CGI can be performed and the image of the target can be directly reconstruced via Eq.(6)\cite{Shapiro2008}. However, in GOAS, the sub-sources are dephased and the relative phase is random and unknown The $m^{th}$ emitted field on the source takes the form of 
	\begin{equation}
		E_0^{~\prime}(\vec \rho_0^{~\prime},t)=\sum_{n=1}^{N}\xi(\vec \rho_0^{~\prime}-\vec \rho_n)e^{i\left[\phi(n,m)+\phi_c(n)\right]}.
		\label{eq:9}
	\end{equation} 
	Here $\phi_c(n)$ is an artificially introduced phase, whose significant role will act in the following. Since the introduced phase in Eq. (8) is different from the relative phase in Eq. (1), the calculated reference patterns will be significantly different from thats imprinted on the target, which is a case far beyond the reach of CGI. Consequently, subjecting Eq. (1)-(5) and Eq. (7)-(8) into Eq. (6), the image of the target cannot be extracted directly, but the spatial information of the target will be involved as (see supplements for details)       
	\begin{equation}
		G(\vec \rho_r)=\alpha \mathcal{T}(\vec \rho_t)e^{-\left|\vec\rho_t \right|^2/a_L^2}*\left| \widetilde{\mathcal{S}}(\vec \rho_r-\vec \rho_t)\right| ^2,
		\label{eq:14}
	\end{equation}  
	where $\alpha$ is constant and $*$ denotes the convolution operation. The GOAS imager of Eq.(9) can be viewed as an active, incoherent imaging system. $\mathcal{T}(\vec \rho_t)$ is the average intensity-reflection coefficient of the target to be imaged. $e^{-\left|\vec\rho_t \right|^2/a_L^2}$ is a Gaussian function with a width of $a_L=L/\omega_0k_0$, corresponding to the field of view. The shift-invariant point-spread function (PSF) of GOAS is dominated by $\widetilde{\mathcal{S}}(\vec \rho_r-\vec \rho_t)$, which is the Fourier transform with a representation as,
	
	\begin{equation}
		\begin{aligned}
			\widetilde{\mathcal{S}}(\vec \rho_r-\vec \rho_t)&=\int_{a_0} d \vec \rho_0 S(\vec \rho_0)e^{i\left[\phi_c(n)-\phi_d(n)-\psi(n)\right]} \\
			&\times e^{ik_0(\vec \rho_r-\vec \rho_t)\vec\rho_0/L},
		\end{aligned}
		\label{eq:14}
	\end{equation}
	in which $S(\vec \rho_0)$ characterizes the distribution of the sub-sources. $\psi(n)$ represents the phase distortion across the source plane induced by turbulence on the source-to-target path. Since both of $\psi(n)$ and $\phi_d(n)$ are random, when the phase $\phi_c(n)$ is arbitrary initially, PSF of GOAS is a random speckle pattern with degeneration caused by turbulence. Hence the initial result $G(\vec \rho_r)$ is featureless and low-contrast. But notably, the PSF possesses an adjustable phase $\phi_c(n)$ thus can work just like deformable mirrors\cite{Roddier}. %Significantly different from traditional adaptive optics, the wavefront correction is performed computationally in GOAS rather than in senser-dependent way.
	%According to the theory of coherence propagation in turbulence\cite{Gbur2001, Salem2003}, the final transverse coherence length of the PSF speckle can be expressed $\rho_L=\max[2L/k_0a_0, 2L/k_0\rho_T]$. That is, when the coherence length of the 
	
	In GOAS, neither the mode number of relative phase or that of the phase distortion induced by turbulence can be larger than the number of the sub-sources\cite{Muller1974}, this therefore allows the adjustable phase $\phi_c(n)$ to computationally compensate the random phase factor in Eq.(10) to near a plane wave. Then the PSF will turn to be a diffraction-limited spot, the width of which is $\rho_L=2L/k_0a_0$. By achieving the computational aberration cancellation, the resolution of GOAS will reach the diffraction limit determined by the baseline, thus we state GOAS is an OAS imager. 
	
	Here we propose to find the optimum compensation phase $\phi_c(n)$ to implement aberration cancellation by maximizing the gradient of $G(\vec \rho_r)$, which is defined as
	
	\begin{equation}
		g=\int d\vec r \left| G(\vec \rho_r)*\Lambda(\vec\rho_r-\vec r)\right| ^2.
		\label{eq:15}
	\end{equation}
	Here  $\Lambda$ is two-dimensional Prewitt operators. In GOAS, a given $\phi_c(n)$ will produce an initial result $G(\vec \rho_r)$ with a certain gradient. Besides, the gradient of $G(\vec \rho_r)$ will increase and reach its maximum when the power of the PSF concentrates from random speckle to a diffraction-limited spot, which corresponds to that the diffraction-limited image is achieved (see supplements for details). Therefore, maximizing the gradient of $G(\vec \rho_r)$ can guide the optimal $\phi_c(n)$ to be found via an iterative process\cite{Conkey2012,Yeminy2021,Hirose2022}.   
	
	% In GOSA, the manually introduced and adjustable $\phi_n^c(\vec \rho_0)$ makes the distributed source act as a deformable mirror in adaptive optics. However, the profound difference is  that wavefornt shaping in GOSA is performed via computational process rather than device-dependent optical modulation.   
	%%In that case, the distribution of reference intensity pattern will maintain the same as that of the actual speckle imprinted on the target, then the GOSA system turn to be a computational GI system and the iamge of the target can be obtained directly. In the follow we will show that $\phi_n^c(\vec \rho_0^{\prime})$ can converge to the optimum, which can compensate both the cophase error and the actual turbulence in the source-to-target path with high fidelity, after an iterative data processing.
	
	%%The image-guided iteration of $\phi_n^c(\vec \rho_0^{\prime})$ can be recognized as a realiztion of computational wavefront shaping process, which corrects the wavefront of the reference field according to the gradient of the image metric $G(\vec \rho_r)$, until the calculated out reference intensity pattern possesses the same distribution of the speckle illuminated on the target and diffraction-limited imaging is achieved. (LCOS-SLM X10468-01)  
	\begin{figure}[h]
		\centering
		\includegraphics[width=\columnwidth]{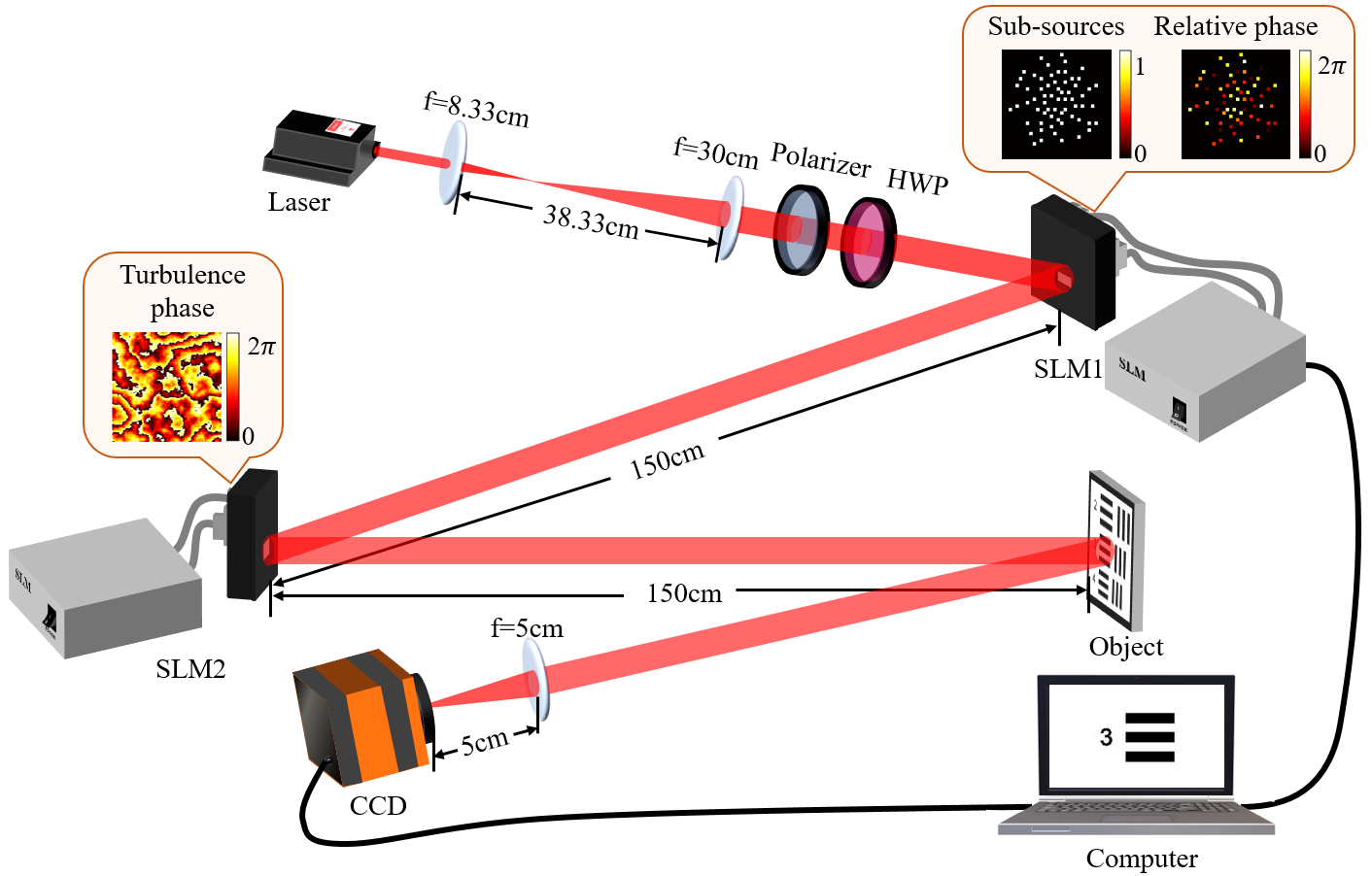}
		\caption{Setup used to demonstrate GOAS. HWP: half wavelength plate. SLM1/SLM2: spatial light modulator(LCOS-SLM X10468-01). The intensity recorded by CCD (Allied Vision Stingray F-125
			B) is integrated spatially as bucket signal. The sub-sources with relative phase are preloaded on SLM1. A Fourier lens or a turbulent phase screen is achieved by SLM2.}
		\label{fig2} 
	\end{figure} 
	
	To demonstrate GOAS imager, a 532nm laser is modulated by 64 isolated macro-pixels (bin 4$\times$4 pixels) with preloaded unknown random phase on SLM1, to achieve 64 sub-sources with random relative phase, as depicted in Fig.2. The size of each macro-pixel is $0.08$mm and the distance between the farthest two macro-pixels is about $2$mm, thus the baseline length is 25 times larger than the beam waist of sub-sources. The light from SLM1 forms speckles, undergoes SLM2 and illuminates the object. Phase screen is induced by SLM2 to simulate the turbulence on the source-to-target path. The distance between the fresh surface of SLM1 and the object is $3$m, thus the spatial resolution determined by the baseline is about $0.80$mm on the object plane. The light reflected back from the object is collected by a CCD, which works as a bucket detector. The reference intensity patterns are calculated via Eq.(7).

	Firstly, GOAS is verified without turbulence. In this case SLM2 work as a Fourier lens with a focus of $1.5$m to meet the far field propagation. The results are shown in Fig.3. When the introduced phase $\phi_c(n)$ is arbitrary, the relative phase cannot be compensated and the PSF is random speckle, causing the initial results featureless as shown in the second row. After the optimum compensation phase is found via the image-guide iterative process, the calculated sequential reference patterns from Eq.(6) will feature the same intensity distribution with the ones imprinted on the object, thus the final images are retrieved\cite{Sun2019,Yang2020} as depicted by the third row. The line width of the strings in Fig. 3d is about $0.89mm$, which is close to the diffraction limit of the synthetic source. In Fig.3n, the lines can be distinguished clearly, implying that the diffraction limit of the GOAS is achieved. 
	\begin{figure}[h]
		\centering
		\includegraphics[width=\columnwidth]{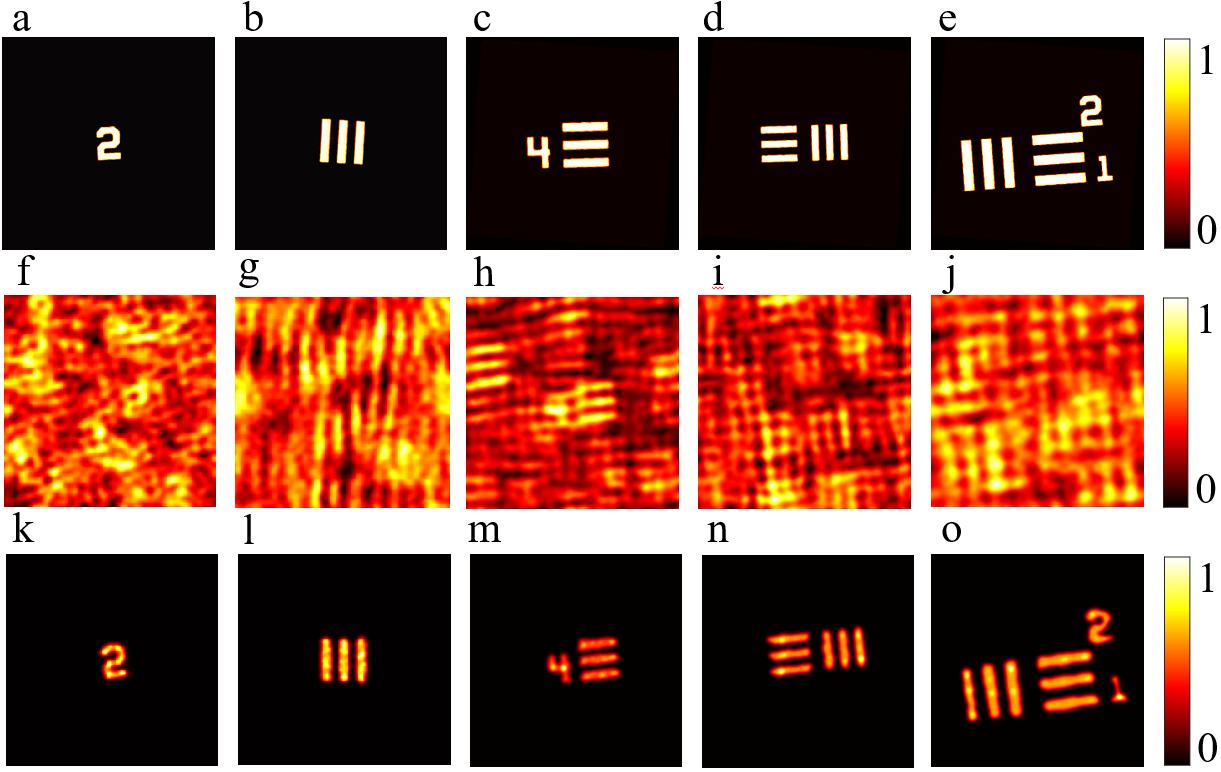}
		\caption{Imaging results without turbulence. a-e: objects from USAF1951. f-g: initial results under the dephased sub-sources. k-o: final results obtained by GOAS after the computational aberration cancellation. All the results are reconstructed with  $4\times10^{4}$ samplings.}
		\label{fig3} 
	\end{figure}
	
	Then the validity of GOAS under turbulence is verified. Fig.4 exhibits the imaging results when SLM2 is loaded a phase screen of turbulence, the strength of which is indicated by the refractive-index structure parameter $C_n^2$ ($m^{-2/3}$)\cite{Lane1992}. The phase screen, which is produced under the a Monte-Carlo method\cite{Schmidt2010}, represents a turbulence volumn with thickness of 300 meters. Since the actual optical path between the synthetic source and the object is $3$m , the geometric size of the experimental setup needs to be spatially scaled up 100 times in the digital
	production the phase screen. This operation can be treated as that the table-top experiment equivalently demonstrate a scene in which the geometric size is expanded by 100 times. Considering the scaling factor, the baseline length of the synthetic source becomes $a_0=0.2$m and the width of the three-slit in Fig.4 is about $0.45$m. In contrast, the transverse coherence length of the turbulence will be $\rho_T=0.08$m or smaller when
	$C_n^2>10^{-14}(m^{-2/3})$, which is calculated as $\rho_T=\left[0.423k_0^2C_n^2\frac{3}{8}L\right]^{-3/5}$\cite{Schmidt2010} with $L$ being the thickness of the turbulence and $L=300$m. That is, regardless of the baseline length of GOAS, the resolution of the system will be degenerated largely when $C_n^2>10^{-14}(m^{-2/3})$. The first and the third rows are the results of CGI\cite{Shapiro2008} with the same distributed sources as GOAS. With the increase of the turbulent strength, deformation and degeneration appear in the images. Especially when $C_n^2>10^{-14}(m^{-2/3})$, the image apparently suffers resolution reduction. As a contrary, original images are still obtained by our method, since the wavefront deformations caused by turbulence are computationally corrected, which is far beyond the reach of CGI.   
	\begin{figure}[h]
		\centering
		\includegraphics[width=\columnwidth]{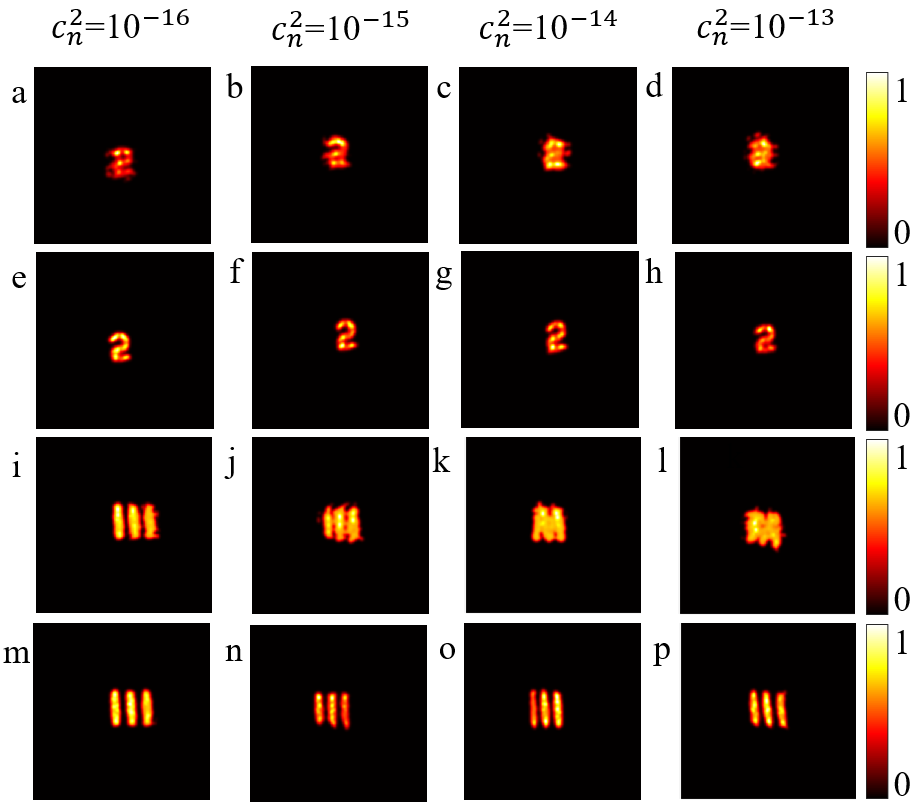}
		\caption{Imaging results under turbulence. a-d and i-l are the reconstructed images from CGI under turbulence with different $C_n^2(m^{-2/3})$; e-h and m-p are the final results of GOSA under the same turbulence. All the results are reconstructed with  $4\times10^{4}$ samplings.}
		\label{fig4} 
	\end{figure}

	In practice, the installation or measurement of the baselines with limited accuracy will cause errors between the actual coordinates of the sub-sources and the one used to calculate the reference intensity pattern, especially when the diameter of whole source reaches meters or larger. Therefore, the robust of GOAS one coordinate error need to be verified. To verify that, coordinate errors are introduced. The actual source realized on SLM1 is depicted as Fig.5(a), while the one used to calculate the reference intensity pattern are shown in the first row of Fig.5. The second and third rows show the retrieved images under different normalized coordinate error, which is quantitated by the ratio between average deviation of the coordinate error and the size of the sub-source as $\sigma_c=\sum_{n=1}^N\left|\vec \rho_0-\vec \rho_0^{~\prime}\right|/N\omega_0$. With the increase of the coordinate error, the mutual coherence degree between the distribution of the actual sub-sources and that of the reference one is reduced, causing the visibility reduction of the retrieved image. However, the reduced mutual coherence is still a peak function when $\sigma_c\sim1$, thus images with considerable quality can still be reconstructed. These results show that installation accuracy that approximates to the beam waist of single sub-source is enough for GOAS. Considering the beam waist of sub-source is several millimeters or larger, this accuracy is not a challenge for current technology. 
	\begin{figure}[h]
		\centering
		\includegraphics[width=\columnwidth]{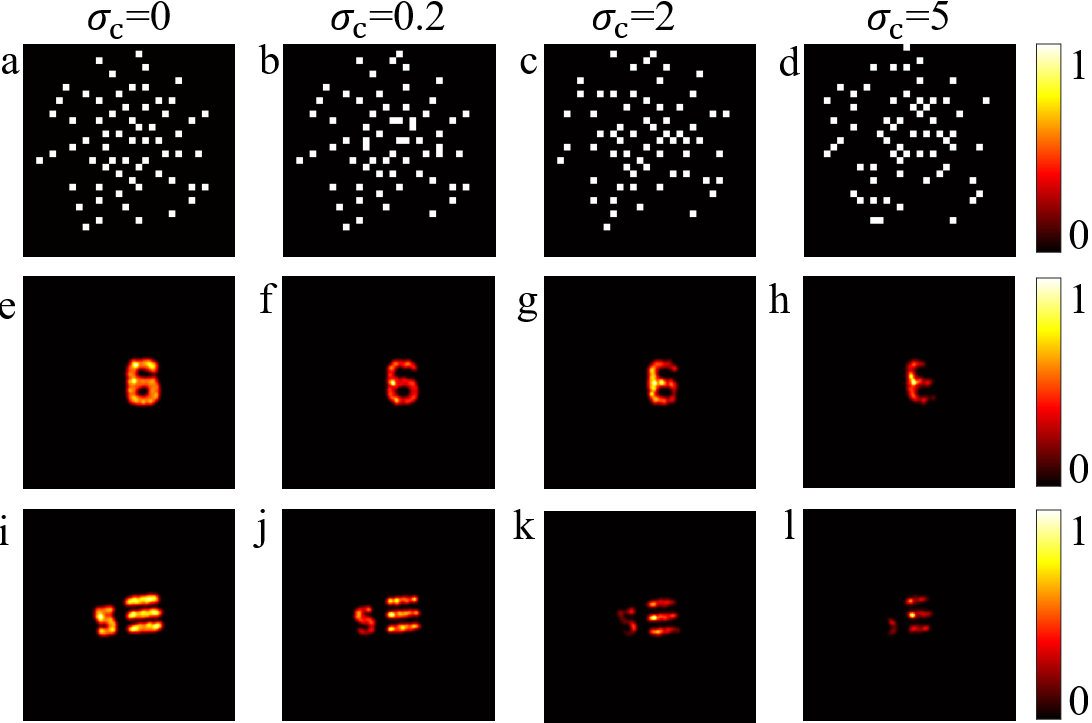}
		\caption{Imaging results under coordinate error. a; the actual distribution of the sub-sources;a-d: the distribution of the sub-sources used to calculate the reference intensity pattern.  e-l: the reconstructed images from GOSA under different coordinate errors. All the results are reconstructed with  $4\times10^{4}$ samplings.}
		\label{fig5} 
	\end{figure} 
	
	Although GOAS is capable to computationally correct the wavefront distortion caused by the weak-to-moderate Kolmogorov-spectrum turbulence, we believe that optical imaging through turbulence with too intense fluctuation is still  a challenge. Compared with the turbulence fluctuating in milliseconds, the relative  phase among the sub-sources usually drifts much slower, especially after the synthetic source being stabilized by means such as vibration isolation, heat insulation and so on. Therefore, once an image of the target has been recovered, the optimum compensation phase $\phi_c(n)$ can be used directly in the following imaging process as the first parents to significantly improve the convergence speed of the iterative process. Compared with the costly calibration technique and devices-based adaptive optics, this new aberration cancellation in GOAS is easy to conduct, and unrestricted by the limited bandwidth, spectral width, dynamic range or other physical parameters of the optical modulation devices.
	
	\section{supplements}
	
	In this supplement, we present the additional details on the theory and experiment of GOAS. The procedure of the image-guide iterative process is also presented in detail.  
	\subsection{the theory of GOAS}
	
	\subsubsection{The synthetic source}
	
	Considering that the length difference among the divided beams is less than the coherence length of the laser pulse but far longer than the wavelength, speckle pattern can be produced after the laser emitted from the distributed, phase-modulated sub-sources. Since the modulation phase of the sub-sources is independent from each other, the temporally varying speckle patterns is pseudo-thermal light when the number of the sub-sources is modest or more\cite{Goodman2007}. Then the cross correlation of the synthetic source can be expressed as
	\begin{equation}
		\begin{aligned}
			\left\langle E_0 (\vec \rho_0,t_1)E_0(\vec \rho_0^{~\prime},t_2)\right\rangle &=\left\langle E_0^{~\prime} (\vec \rho_0,t_1)E_0^{~\prime}(\vec \rho_0^{~\prime},t_2)\right\rangle\\
			&=\left\langle E_0 (\vec \rho_0,t_1)E_0^{~\prime}(\vec \rho_0^{~\prime},t_2)\right\rangle\\
			&=0,         \\
			\left\langle E_0 (\vec \rho_0,t_1)E_0^*(\vec \rho_0^{~\prime},t_2) \right\rangle &=\left\langle E_0^{~\prime*} (\vec \rho_0,t_1)E_0^{~\prime}(\vec \rho_0^{~\prime},t_2)\right\rangle   \\
			&= \left\langle E_0^* (\vec \rho_0,t_1)E_0^{~\prime}(\vec \rho_0^{~\prime},t_2) \right\rangle\\
			&=\frac{2NP}{\pi \omega_0^2}S(\vec \rho_0)e^{-\frac{|\vec \rho_0-\vec \rho_0^{~\prime}|^2}{2\omega_0^2}}
			e^{-\frac{|t_1-t_2|^2}{2T_0^2}},
		\end{aligned}
		\label{eq:1}
		\tag{S1}
	\end{equation}
	Here $\left\langle \cdot \right\rangle $ represents the ensemble average for classical correlation. $S(\vec \rho_0)=\sum_{n=1}^Ne^{-\frac{\left|\vec \rho_0-\vec \rho_n\right |^2}{2\omega_0^2}}$, characterizes the intensity distribution of the sub-sources.  $S(\vec \rho_0)$ is a two-dimensional, broaden (the width is determined by the waist of the sub-source) comb function with random distribution, which accounts for the dozens of randomly, widely distributed sub-sources. To ensure that the sub-apertures are approximately nonoverlapping, the center coordinate and the beam waist of the sub-source subject to $\min_{k\neq j}|\vec\rho_k-\vec\rho_j|>3\omega_0$. $P$ is the intensity of each sub-source with unit $Watts$. $T_0$ is the coherence time of the light field. In Fig.1, the temporally varying speckle pattern is produced by randomly modulating the phase of each sub-source, thus $T_0$ is the reciprocal of the bandwidth of the phase modulators. If a bucket detector with a higher bandwidth is used, which means the detector can easily record the signal within $T_0$, then the factor of coherence time in the correlation results can be ignored. In GOAS, each sub-source also gets a relative phase $\phi_d(n)$ on the signal path while an artificially introduced phase $\phi_c(n)$ when calculate the reference pattern. Since both of the phases are treated as constant and independent of the ensemble average, the phase-insensitive cross correlation of the source can be written as 
	
	\begin{equation}
		\left\langle E_0^* (\vec \rho_0,m)E_0^{~\prime}(\vec \rho_0^{~\prime},m) \right\rangle\\=\frac{2NP}{\pi \omega_0^2}S(\vec \rho_0)e^{-\frac{|\vec \rho_0-\vec \rho_0^{~\prime}|^2}{2\omega_0^2}}e^{ i[\phi_c(n)-\phi_d(n)] }
		\label{eq:1}
		\tag{S2}
	\end{equation}
	where $m$ is the index of emitted illumination pattern and the complex phase factor $e^{ i[\phi_c(n)-\phi_d(n)] }$ accounts for its invariance over time. Considering the profile of  $S(\vec \rho_0)$, the synthetic source can also be seen as that a pseudo-thermal source with diameter $a_0$ is masked except for $N$ small sub-sources with waist of $\omega_0$, satisfying $\omega_0\ll a_0$. Besides, the relative phase among the sub-sources is random and unkown.  %This synthetic source can not be used to directly perform traditional CGI, since which requires the whole profile of the light field on the source is known. %Notably, the structure of the synthetic source here is samiliar with a famous set-up, called a Fizeau interferometer\cite{}, in which a telescope is masked in the way here. In the following, we will demonstrate the image formation is like the  %In practice, the diameter of each sub-sources can be several millimeters or smaller, but the length of the baseline can be tens of meters or larger.}
	
	\subsubsection{propagation through turbulence}
	The phase-modulated laser emitted from the synthetic source propagates through  $L$-m turbulence and illuminates the object.  The light reflected by the target undergoes $L$-m turbulence and arrives the plane of bucket detector. Because the turbulence is random, they can only be described statistically by using statistical estimates such as variances, or covariances. As an example, the mutual coherence function of the turbulence on the source-to-target and target-to-detector path follows\cite{Hardy2011}
	\begin{equation}
		\begin{aligned}
			&\left<e^{\Psi_t^*(\vec \rho_t,\vec \rho_0)} e^{\Psi_d^(\vec \rho_t^{~\prime}, \vec \rho_0^{~\prime})}\right>\\
			&~~=e^{-(|\vec \rho_t-\vec \rho_t^{~\prime}|^2+(\vec \rho_t-\vec \rho_t^{~\prime})\cdot(\vec \rho_0-\vec \rho_0^{~\prime})+|\vec \rho_0-\vec \rho_0^{~\prime}|^2)/2\rho_T^2}
		\end{aligned}
		\label{eq:3}
		\tag{S3}
	\end{equation} 
	and
	\begin{equation}
		\begin{aligned}
			&\left<e^{\Psi_d^*(\vec \rho_d,\vec \rho_t)} e^{\Psi_d^(\vec \rho_d^{~\prime}, \vec \rho_t^{~\prime})}\right>\\
			&~~=e^{-(|\vec \rho_t-\vec \rho_t^{~\prime}|^2+(\vec \rho_t-\vec \rho_t^{~\prime})\cdot(\vec \rho_d-\vec \rho_d^{~\prime})+|\vec \rho_d-\vec \rho_d^{~\prime}|^2)/2\rho_T^2}
		\end{aligned}
		\label{eq:4}
		\tag{S4}
	\end{equation} 
	respectively.
	Here $\rho_T$ is the spatial coherence length of the turbulence and $\rho_T=\left[0.423k_0^2C_n^2\frac{3}{8}L\right]^{-3/5}$ for plane waves\cite{Schmidt2010}. $C_n^2$ is the refractive index structure parameter of the turbulence in the propagation direction\cite{Lane1992}, which can be treated approximately as a constant when the angle between the propagation direction and the horizontal is small. %an interesting property of $\rho_T$ is that the root mean square (rms) phase distortion over a circular area of diameter $\rho_T$ is about 1 radian.
	
	\subsubsection{target and bucket signal}
	In the natural world, almost all the objects possess a rough surface with random depths substantially exceeding the optical wavelength\cite{Goodman2007}. Under the illumination of speckle with a small transverse coherence length, the rough-surfaced target can be treated as a quasi-Lambertian reflector with a random field-reflection coefficient $T(\vec \rho_t)$\cite{Shapiro1981}. Following Lidar theory\cite{Shapiro1981,Hardy2011,Hardy2013},  the target possesses the autocorrelation function,
	\begin{equation}
		\left<T^*(\vec \rho_t)T(\vec \rho_t^{~\prime})\right>=\lambda^2\mathcal{T}(\vec \rho_t)\delta(\vec \rho_t-\vec \rho_t^{~\prime})
		\label{eq:5}
		\tag{S5}
	\end{equation}
	where $\delta(\cdot)$ is a Drac function, $\mathcal{T}(\vec \rho_t)$ is the average intensity-reflection coefficient and $\mathcal{T}(\vec \rho_t)\leq 1$. 
	
	Under the illumination of $M$ temporally varying speckle patterns, the $m^{th}$ reflected field from the target is collected by the bucket detector. The bucket signal can be evolved from Eq. (5) as
	\begin{equation}
		i_b(m)= \frac{q\eta A_b }{hf_0}\left\langle E_d^*(\vec\rho_d,m) E_d(\vec\rho_d,m)\right\rangle
		\label{eq:7}
		\tag{S6}
	\end{equation}  
	in which $A_b\equiv\int d \vec \rho_d\mathcal{A}_b^2(\vec \rho_d)$ represents the area of the photosensitive region of the bucket detector.

	\subsubsection{initial result in GOAS}
	In GOAS,  the initial result $G(\vec \rho_r)$ is calculated from the cross correlation between the bucket signals and the reference intensity pattern shown by Eq.(6). In the following, we will conduct the derivation of the initial results. 
	
	%due to the unkown relative phase among the sub-sources, the initially calculated reference pattern is features different intensity distributions from the real one imprinted on the object, thus the image of the target can not be reconstructed directly.
	According to Eq.(S6), the first term in the right of Eq.(6) can be written as
	\begin{equation}
		\begin{aligned}
			&\left < i_b(m)I(\vec \rho_r,m)\right >\\
			&= \frac{q\eta A_b}{hf_0}  \left< E_d^*(\vec \rho_d,m)E_r^*(\vec \rho_r,m)E_r(\vec \rho_r,m)E_d(\vec \rho_d,m)\right>
		\end{aligned}
		\label{eq:10}
		\tag{S7}
	\end{equation} 
	where the second-order moment of intensity is evolved into the fourth-order moment of the field arriving the bucket detector and the reference one. When the back-propagate from the bucket detector to the target\cite{Hardy2011,Hardy2013} is used, 
	\begin{widetext}
		\begin{equation}
			\begin{aligned}
				\left< E_d^*(\vec \rho_d, m)E_r^*(\vec \rho_r, m)E_r(\vec \rho_r, m)E_d(\vec \rho_d, m)\right >&=\int d \vec \rho_t\int d \vec \rho_t^{~\prime}\left<T^*(\vec \rho_t)T(\vec \rho_t^{~\prime})\right>\left\langle e^{\Psi_d^*(\vec \rho_d,\vec \rho_t)}e^{\Psi_d(\vec \rho_d,\vec \rho_t^{~\prime})}\right\rangle \\
				&\times\left< E_t^*(\vec \rho_t, m)E_r^*(\vec \rho_r, m)E_r(\vec \rho_r, m)E_d(\vec \rho_t^{~\prime}, m)\right >\\
				&\times\frac{k_0^2e^{ik_0(|\vec \rho_t^{~\prime}|^2-|\vec \rho_t|^2)/2L+ik_0\vec \rho_d(\vec \rho_t-\vec \rho_t^{~\prime})/L}}{4\pi^2L^2}
			\end{aligned}
			\label{eq:11}
			\tag{S8}
		\end{equation}
	\end{widetext}
	Considering the statistical property of the turbulence shown by Eq.(S4) and that of the target's reflection coefficient expressed as Eq.(S5), Eq.(S8) can be simplified as 
	\begin{equation}
		\begin{aligned}
			& \left< E_d^*(\vec \rho_d, m)E_r^*(\vec \rho_r, m)E_r(\vec \rho_r, m)E_d(\vec \rho_d, m)\right >\\
			& =\frac{1}{L^2}\int d \vec \rho_t\mathcal{T}(\vec \rho_t)\times \left< E_t^*(\vec \rho_t, m)E_r^*(\vec \rho_r, m)E_r(\vec \rho_r,t)E_t(\vec \rho_t, m)\right >
		\end{aligned}
		\label{eq:13}
		\tag{S9}
	\end{equation}
	which exhibits that the fourth-order moment of the field arriving on the bucket detector and the reference one is proportional to the fourth-order moment of the field imprinted on the target and the reference one. That is, the turbulence on the target-to-detector path will not affect the cross correlation results. This arises from the statistically-independent randomness of the turbulence and the that of the quasi-Lambertian target's surface. In addition, when the target is not quasi-Lambertian, Eq.(S9) can be still reasonable under the condition that the diameter of the lens in front of the bucket detector is large enough that the intensity fluctuation of the reflected light from the target induced by the turbulence is negligible. This condition is not difficult to achieve, since the lens in front of the bucket detector is to concentrate the energy rather than imaging. Energy-concentration lens is less unaffected by aberrations, thus is much easier to obtain than manufacturing an imaging lens with the same size.   
	
	According to the isotropy of the Kolmogorov-spectrum turbulence, the light field imprinted on the target plane $E_t(\vec \rho_t, m)$ can still be treated as obeying Gaussian random distribution, so does $E_r(\vec \rho_r, m)$. Then the Complex Gaussian moment theorem can be used and the fourth-order field moment in Eq. (S9) can be expressed as\cite{Goodman2007}
	\begin{widetext}
		\begin{equation}
			\begin{aligned}
				\left< E_t^*(\vec \rho_t, m)E_r^*(\vec \rho_r, m)E_r(\vec \rho_r, m)E_d(\vec \rho_t,m)\right >&= \left< E_r^*(\vec \rho_r, m)E_r(\vec \rho_r, m)\right >\left< E_t^*(\vec \rho_t, m)E_t(\vec \rho_t, m)\right >+\left|\left\langle E_t^*(\vec \rho_t, m)E_r(\vec \rho_r, m)\right\rangle \right|^2\\
			\end{aligned}
			\label{eq:11}
			\tag{S10}
		\end{equation}
	\end{widetext}
	where the first term of the right hand is the product between the average intensity of the reference pattern and that of the pattern imprinted on the target. The second term is the square of the first-order mutual coherence of the reference field and the field imprinted on the target. 
	Substituting Eq.(S7), (S9) and (S10)to Eq.(6), we have 
	\begin{widetext}
		\begin{equation}
			\begin{aligned}
				G(\vec \rho_r)& = \frac{q\eta A_b}{hf_0L^2}\int d \vec \rho_t \mathcal{T}(\vec \rho_t)\left[\left<I(\vec \rho_t, m)\right>\left<I(\vec \rho_r, m)\right>+\left|\left<E_t^*(\vec \rho_t, m)E_r(\vec \rho_r, m)\right>\right|^2\right]-\left < i_b( m)\right >\left < I(\vec \rho_r, m)\right >\\
				&=\left < i_b( m)\right >\left < I(\vec \rho_r, m)\right >+\frac{ q\eta A_b}{hf_0L^2}\int d \vec \rho_t \mathcal{T}(\vec \rho_t)\left|\left<E_t^*(\vec \rho_t, m)E_r(\vec \rho_r, m)\right>\right|^2-\left < i_b( m)\right >\left < I(\vec \rho_r, m)\right >\\
				&=\frac{ q\eta A_b}{hf_0L^2}\int d \vec \rho_t \mathcal{T}(\vec \rho_t)\left|\left<E_t^*(\vec \rho_t, m)E_r(\vec \rho_r, m)\right>\right|^2
			\end{aligned}
			\label{eq:12}
			\tag{S11}
		\end{equation}
	\end{widetext}
	in which the product of the average intensity is eliminated and the square of the first-order mutual coherence is left in the integral. As a integral kernel of the image formation, the square of the first-order mutual coherence plays an important role to characterize the features of GOAS (we will detail below). Here we discuss the case that the path length $L$ satisfies far field propagation of the synthetic pseudo-thermal source, which means $k_0a_0\omega_0/2L\ll 1$ \cite{Erkmen2008,Erkmen2010,Shapiro2012}. Then the second-order moments of cross correlation in Eq.(S11) can be written as the coherence propagation of the mutual coherence of the source, as
	\begin{widetext}
		\begin{equation}
			\begin{aligned}
				\left<E_t^*(\vec \rho_t, m)E_r(\vec \rho_r, m)\right>&=\int_{a_0} d \vec \rho_0 \int_{a_0} d \vec \rho_0^{~\prime}\left<E_0^*(\vec \rho_0, m)E_0^{~\prime}(\vec \rho_0^{~\prime}, m)\right>e^{\left\lbrace \chi(\vec \rho_t,\vec \rho_0)-i\psi(\vec \rho_t,\vec \rho_0)\right\rbrace }\frac{k_0^2e^{ik_0(|\vec \rho_r|^2-|\vec \rho_t|^2)/2L+(\vec \rho_r\cdot\vec \rho_0^{~\prime}-\vec \rho_t\cdot\vec\rho_0)/L}}{4\pi^2L^2}\\
				&\approx \int_{a_0} d \vec \rho_0 \int_{a_0} d \vec \rho_0^{~\prime}\left<E_0^*(\vec \rho_0, m)E_0^{~\prime}(\vec \rho_0^{~\prime}, m)\right>e^{-i\psi(\vec \rho_t,\vec \rho_0) }\frac{k_0^2e^{ik_0(|\vec \rho_r|^2-|\vec \rho_t|^2)/2L+(\vec \rho_r\cdot\vec \rho_0^{~\prime}-\vec \rho_t\cdot\vec\rho_0)/L}}{4\pi^2L^2},		
			\end{aligned}
			\label{eq:13}
			\tag{S12}
		\end{equation}
	\end{widetext}
	where $\chi(\vec \rho_t,\vec \rho_0)$ and $\psi(\vec \rho_t,\vec \rho_0)$ represents the log amplitude and phase fluctuations of the field caused by the turbulence on source-to-target path respectively, and $\Psi_t^*(\vec \rho_t,\vec \rho_0)=\chi(\vec \rho_t,\vec \rho_0)-i\psi(\vec \rho_t,\vec \rho_0)$. The approximation arises from that image quality can be degraded by both phase and amplitude distortions of the optical wavefront but the effect of phase fluctuations is predominant\cite{Roddier,Dixon2011,Chan2011}.  Substituting Eq.(S2) into Eq.(S12), one can obtain 
	\begin{widetext}
		\begin{equation}
			\begin{aligned}
				\left<E_t^*(\vec \rho_t, m)E_r(\vec \rho_r, m)\right>&=\frac{NPk_0^2e^{ik_0(|\vec \rho_r|^2-|\vec \rho_t|^2)/2L}}{2\pi^3L^2 \omega_0^2}\int_{a_0} d \vec \rho_0 \int_{a_0}d \vec \rho_0^{~\prime}S(\vec \rho_0)e^{i\left[\phi_c(n)-\phi_d(n)\right]}e^{-i\psi(\vec \rho_t,\vec \rho_0)}e^{-\frac{|\vec \rho_0-\vec \rho_0^{~\prime}|^2}{2\omega_0^2}}e^{ik_0(\vec \rho_r \cdot\vec\rho_0\cdot-\vec \rho_t\cdot\vec\rho_0^{~\prime})/L} \\
				&=\frac{NPk_0^2e^{ik_0(|\vec \rho_r|^2-|\vec \rho_t|^2)/2L}}{2\pi^3L^2 \omega_0^2}\int_{a_0} d \vec \rho_0 S(\vec \rho_0)e^{i\left[\phi_c(n)-\phi_d(n)-\psi(\vec \rho_t,\vec \rho_0)\right]} e^{ik_0(\vec \rho_r-\vec \rho_t)\vec\rho_0/L}\int_{\omega_0}d \Delta\vec \rho e^{-\frac{\left| \Delta\vec \rho\right| ^2}{2\omega_0^2}}e^{ik_0\vec \rho_t \Delta\vec \rho/L}\\
				&=\frac{NPk_0^2e^{ik_0(|\vec \rho_r|^2-|\vec \rho_t|^2)/2L}}{2\pi^3L^2 \omega_0^2}e^{-\omega_0^2k_0^2\left|\vec\rho_t \right|^2/2L^2 }\int_{a_0} d \vec \rho_0 S(\vec \rho_0)e^{i\left[\phi_c(n)-\phi_d(n)-\psi(n)\right]} e^{ik_0(\vec \rho_r-\vec \rho_t)\vec\rho_0/L}\\
				&=\frac{NPk_0^2e^{ik_0(|\vec \rho_r|^2-|\vec \rho_t|^2)/2L}}{2\pi^3L^2 \omega_0^2} e^{-\left|\vec\rho_t \right|^2/2a_L^2}\widetilde{\mathcal{S}}(\vec \rho_r-\vec \rho_t)
			\end{aligned}
			\label{eq:13}
			\tag{S13}
		\end{equation}
	\end{widetext}
	%For this reason, fluctuations of the wave-front amplitude contribute much less to image degradation than the wave-front phase. However, if the wave-front phase is compensated by an AO system, there will be a residual image degradation due to the fluctuations of the wave-front amplitude. The loss in Strehl ratio is only a few percent in the infrared but may reach 10–15\% at visible wavelengths (Roddier and Roddier 1986)
	in which $\Delta \vec \rho=\vec\rho_0-\vec\rho_0^{~\prime}$. $a_L=L/\omega_0k_0$, is the beam waist of the sub-source on the target plane, which is also the field of view (FOV) illuminated by the source. $\widetilde{\mathcal{S}}(\vec \rho_r-\vec \rho_t)$ is a Fourier transform with a representation as,
	\begin{equation}
		\begin{aligned}
			\widetilde{\mathcal{S}}(\vec \rho_r-\vec \rho_t)&=\int_{a_0} d \vec \rho_0 S(\vec \rho_0)e^{i\left[\phi_c(n)-\phi_d(n)-\psi(n)\right]} \\
			&\times e^{ik_0(\vec \rho_r-\vec \rho_t)\vec\rho_0/L},
		\end{aligned}
		\label{eq:14}
		\tag{S14}
	\end{equation}
	in which the transform kernel is a complex-valued function. The real part of the kernel is the distribution of the synthetic source, which determines the spatial frequency of the PSF. The imaginary part is three phase factors, which includes the adjustable compemsation phase  $\phi_c(n)$, the unknown relative phase  $\phi_d(n)$ and phase fluctuation  $\psi(n)$ induced by turbulence. Notably, the spatial arguments of phase fluctuation $\psi(\vec \rho_t,\vec \rho_0)$ is replaced by  $\psi(n)$ in Eq. (S13). This means the wavefront distortion caused by the turbulence on source-to-target path is represented by a phase screen  $\psi(n)$ and be corrected on the source plane\cite{Tyson2022,Gruneisen2021}. In our method, the spatial frequency of the PSF is optimized by designing both the number and the spatial distribution of the sub-sources thus wavefront distortion can be computationally corrected well. In principle, except for source plane, the wavefront distortion can also be computationally corrected by inserting a phase screen on another plane away from the source when calculating the reference pattern. The degeree of the inserted phase screen and the distance between the phase screen and the source will the ifluence on the finally FOV of imaging, dut to the anisoplanicity of turbulence\cite{Roddier,fried1982}.

	Substituting  Eq.(S13) into Eq.(S11), the initial results from the cross correlation between the bucket signal and the reference pattern becomes 
	\begin{equation}
		\begin{aligned}
			G(\vec \rho_r)&=\frac{4 q\eta A_bN^2P^2}{hf_0\omega_0^4\lambda^4L^6}\int d \vec \rho_t \mathcal{T}(\vec \rho_t)e^{-\left|\vec\rho_t \right|^2/a_L^2}\left| \widetilde{\mathcal{S}}(\vec \rho_r-\vec \rho_t)\right| ^2\\
			&=\alpha \mathcal{T}(\vec \rho_t)e^{-\left|\vec\rho_t \right|^2/a_L^2}*\left| \widetilde{\mathcal{S}}(\vec \rho_r-\vec \rho_t)\right| ^2
		\end{aligned}
		\label{eq:12}
		\tag{S15}
	\end{equation}
	in which  $\alpha=\frac{4 q\eta A_bN^2P^2}{hf_0\omega_0^4\lambda^4L^6}$ is constant. Eq. (S15) expresses an incoherent imaging system. The $\mathcal{T}(\vec \rho_t)$ is the target to be imaged. $e^{-\left|\vec\rho_t \right|^2/a_L^2}$ is a Gaussian function with a width of $a_L=L/\omega_0k_0$, which actually determines the largest FOV. The shift-invariant PSF of GOAS is characterized by $\left| \widetilde{\mathcal{S}}(\vec \rho_r-\vec \rho_t)\right|^2$. As shown by Eq. (S14), the PSF with adjustable phase  $\phi_c(n)$ can work just like deformable mirrors to compensate the relative phase and the turbulent distortion. Significantly different from traditional adaptive optics, the wavefront correction is performed computationally in GOAS rather than in senser-dependent way. The challenge is to find out the optimum  $\phi_c(n)$, which will explained in the next part.%when the artificially introduced phase  $\phi_c(n)$ is arbitrary initially, the PSF is a random speckle pattern with the degeneration caused by the turbulence, resulting in the result $G(\vec \rho_r)$ featureless and low-contrast. However, if the phase factor can be compensated as a plane-like phase, $\widetilde{\mathcal{S}}(\vec \rho_r-\vec \rho_t)$ will turn to be a spot with a width of $\rho_L=2L/k_0a_0$. Then, the resolution of GOAS will reach the diffraction limit determined by the baseline length of the synthetic source, which is far larger than the size of sub-source.
	
	%width of the ψm(ρ,ρ) is a complex-valued random process that encapsulates the effects of turbulence
	\subsection{The details of the GOAS experiment} 
	\subsubsection{image-guide iterative process} 
	% the sampling process means projecting a sequence of random speckle patterns and recording the resultant bucket signals. The sequential random pattern is produced by the temporally varying modulation phase plus the relative phase imposed on each sub-source. In the computation of reference speckle pattern, the modulation phase plus a compensation phase $\phi_c(\vec \rho_0)$ is used. That is, a fixed $\phi_c(\vec \rho_0)$ is occupied to calculate the sequential reference pattern via Eq. (7). With the sequential reference pattern and the sequence of bucket signal, an initial result $G(\vec \rho_r)$ can be reconstructed via Eq.(6). Therefore, a given $\phi_c(\vec \rho_0)$ will produce an image results from Eq. (6) and the image will possess a certain contrast. Initially, the arbitrary introduced phase $\phi_c(\vec\rho_0)$ fails to compensate the relative phase and the phase distortion, the PSF of GOSA will be a random speckle and the contrast of  $G(\vec r_r)$ will be very low. With both of the relative phase and the phase distortion is partially compensated by  $\phi_c(\vec r_0)$, the power of the PSF gradually concentrate from random speckle to a peak function with some background noise, and thereby the contrast of calculated  $G(\vec r_r)$ will increase. Ideally, if the relative phase and the phase distortion is perfectly compensated, the PSF will turn to be a diffraction-limited spot and clear image of the target will be obtained. Therefore,
	In GOAS, image sharpening can be used to guide the the optimization of the compensation phase   $\phi_c(n)$ to be found adaptively\cite{Muller1974}. To achive that, the sharpness of the initial result can be quantified by the gradient as
	\begin{equation}
		g=\sum_x\sum_y\left[(G(x,y)*\Lambda_x)^2+(G(x,y)*\Lambda_y)^2\right].
		\label{eq:15}
		\tag{S16}
	\end{equation}
	where * represents the convolution operation, $x$ and $y$ are two orthogonal component of the vector $\vec\rho_r$, $\Lambda_x$ and $\Lambda_y$ are Prewitt operators at two directions and 
	\begin{equation}
		\Lambda_x=\left[                 
		\begin{array}{ccc}   
			-1 & -1 & -1\\  
			0 & 0 & 0\\
			1  & 1 & 1\\ 
		\end{array}
		\right],
		\Lambda_y=\left[                 
		\begin{array}{ccc}   
			-1 & 0 & 1\\  
			-1 & 0 & 1\\
			-1 & 0 & 1\\ 
		\end{array}
		\right].               
		\label{eq:15}
		\tag{S17}
	\end{equation}

	\begin{figure}[h]
		\centering
		\includegraphics[width=\columnwidth]{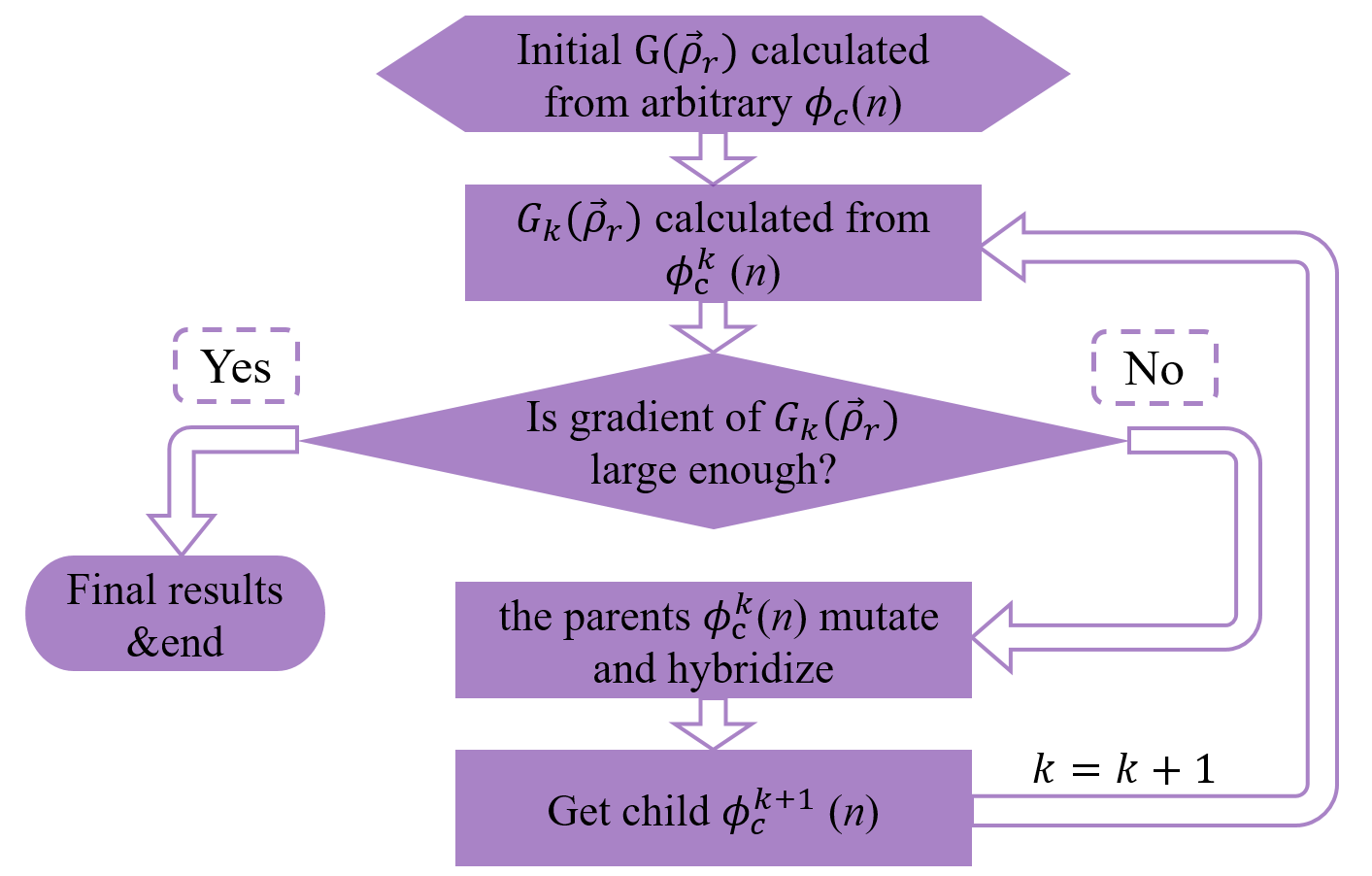}
		\caption{Flow diagram of the image-guide iterative process to fing out the optimum compensation phase.}
		\label{figs1} 
	\end{figure} 
	Eq.(S16) is the discretization of Eq.(11) for the convenient of data processing.
	
	Fig. S1 shows the flow diagram of the image-guide iterative process. Initially, when an arbitrary  $\phi_c(n)$ is used, the contrast of $G(\vec \rho_r)$ is low and the  gradient calculated via Eq. (S16) is small. Then the algorithm starts to iterative, each cycle of which include three steps. Firstly, the parent population of  $\phi_c^k(n)$ is generated and the corresponding $G_k(\vec \rho_r)$ can be obtained. Secondly, mutate the subpopulation of  $\phi_c^k(n)$ which can produce $G_k(\vec \rho_r)$ with larger gradient. Thirdly hybridize the mutated subpopulation with another new random population of  $\phi_c^k(n)$ to give birth to the child population  $\phi_c^{k+1}(n)$. The child population is also the parent in the next cycle, until the gradient of $G_k(\vec \rho_r)$ is large enough. In this kind of image-guide iterative process, the guide functions are usually independent with the position of the image, thus the lateral position of the object will be discarded in the final image. 
	
	\begin{figure}[h]
		\centering
		\includegraphics[width=\columnwidth]{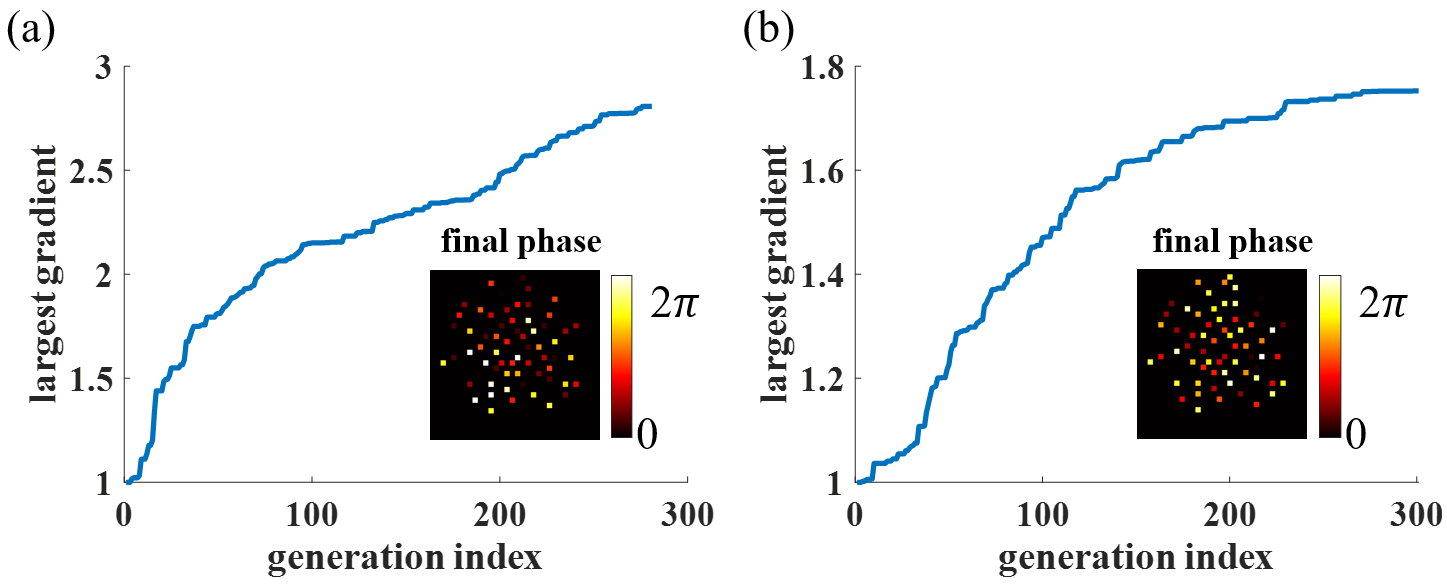}
		\caption{The largest gradient of $G_k(\vec \rho_r)$ of every generation. Y-axis is the largest gradient among all the $G_k(\vec \rho_r)$ produced by $k$th population of  $\phi_c^k(n)$}. 
		\label{figs2} 
	\end{figure} 
	
	In our experiment, 4$\times10^4$ samplings are performed to retrieve the image. That means with each  $\phi_c^k(n)$, 4$\times10^4$ reference pattern and the subsequent result $G_k(\vec\rho_r)$ are calculated. The number of the sub-sources is 64. The population size in each cycle is 128 and the initial mutation rate is 30\%, which will be reduced with the increase of the generation index. Once the mean error of the compensated phase is smaller than $\pi/2$, an image $G_k(\vec \rho_r)$ with considerate contrast can be obtained. Consequently, the gradient of the image converges quickly under the iterative process. Fig. S2(a) and S2(b) shows the largest gradient of every population in the iterative process to obtain Fig. 3(o) and Fig.4(g), respectively. The both gradient of the $G_k(\vec \rho_r)$ start to converge when $k\sim300$. If more sub-sources are used, the phase distortion caused by the turbulence can be compensated more accurately and an image with higher contrast can be obtained. In that case, the converges speed of the image-guide iterative process will not be reduced largely if more RAM can be used and the parallelism of data processing can be improved.    
	
	\subsubsection{The turbulence phase screen} 
	%In GOAS, if the sampling process is finished within the coherence time of the turbulence, the time consumption of the data processing will not reduce the quality of the image. 
	The turbulent phase screen preloaded on SLM2 is produced via a Monte-Carlo method \cite{Schmidt2010}. A phase screen represents an extended turbulent volume, and the parameter of the produced phase screen is closely related to the thickness of the turbulence, which is usually $100m$ or thicker. However, the distance between the source and the target in our table-top experiment is only $3m$, which is too short for a practical phase screen. To meet the geometry required by phase screen model, the actual size of the experimental setup is spatially scaled up 100 times in the digital production the phase screen. That is, the lab-table experiment equivalently demonstrate a scene where the distance between the source and the object is $300m$. Consequently, considering the scaling factor, the baseline length of the synthetic source is enlarged as $a_0=0.2m$, the beam waist of the sub-source is scaled up to $0.008m$ and the size of the object becomes about $0.45m$.
	
	In traditional adaptive optics, the compensated FOV is limited after wavefront correction. It occurs because of the differences between wavefront coming from different directions, which is called the anisoplanicity of the turbulence\cite{fried1982}. If a guide source is used to sense the wavefront, the compensation will be good only for objects close enough to the guide source. As the angular distance between the object and the guide source increases, image quality decreases. When one can accept Mean Square Error (MSE) of the compensation less than 1 radian, the FOV $\theta$ subjects to $\theta\propto\frac{\rho_T}{d}$, in which $\rho_T$ is the coherence length of the turbulence and $d$ is the distance between the turbulence layer and the wavefront compensation devices. In GOAS, an image with considerate contrast can be obtained once the MSE of the compensation is less than $\pi/2$, thus the FOV of GOAS can be larger than traditional adaptive optics. Better yet, GOAS allows to compensate the turbulence in the source-to-object path by inserting a phase screen on the conjugate plane in the reference path, which means the distance between the turbulence-layer and the source is almost equal to the distance between the correction plane and the source. Then $d$ can next to $zero$ and the FOV $\theta$ of GOAS will be large enough.

	\section{acknowledgments}
	
	This work is supported by the National Natural Science Foundation of China under Grant Nos. 62105365 and 62275270.%Research Program of National University of Defense Technology (ZK22-58)

\end{CJK}

\section{Reference}
	\begin{thebibliography}{35}
		
		
		
		\bibitem{Johnson1974}
		M.Johnson, A. Betz, and C. H. Townes, 10-μm heterodyne stellar interferometer, Phys. Rev. Lett. 33(27), 1617 (1974).
		\bibitem{Labeyrie2006}
		A. Labeyrie, S. Lipson, P. Nisenson, An introduction to optical stellar interferometry[M]. Cambridge University Press, (2006).
		\bibitem{Saha2002}
		S. Saha, Modern optical astronomy: technology and impact of interferometry, Rev. Mod. Phys., 74(2), 551 (2002).
		\bibitem{HBT1956}
		B. Hanbury and R. Twiss, A test of a new type of stellar interferometer on Sirius, Nature, 178, 1046-1048 (1956).
		\bibitem{Boal1990}
		D. Boal, C. Gelbke, and B. Jennings, Intensity interferometry in subatomic physics, Rev. Mod. Phys. 62(3), 553 (1990).
		\bibitem{Yabashi2001}
		M. Yabashi, K. Tamasaku, and T. Ishikawa, Characterization of the transverse coherence of hard synchrotron radiation by intensity interferometry, Phys. Rev. Lett.  87(14), 140801 (2001).
		\bibitem{Labeyrie1999}
		A. Labeyrie, Snapshots of Alien Worlds--The Future of Interferometry, Science, 285(5435), 1864-1865 (1999).
		\bibitem{Moreira2013}
		A. Moreira, P. Prats-Iraola, M. Younis, and et al, A tutorial on synthetic aperture radar, IEEE Geosc. Rem. Sen. M , 1(1), 6-43 (2013).
		\bibitem{Giovannetti2001}
		V. Giovannetti, S. Lloyd, L. Maccone, and F. Wong,  Clock synchronization with dispersion cancellation, Phys. Rev. Lett. 87(11), 117902 (2001).
		\bibitem{Roggemann1997}
		M. Roggemann, B. Welsh, and R. Fugate,  Improving the resolution of ground-based telescopes, Rev. Mod. Phys. 69(2), 437 (1997).
		\bibitem{Gruneisen2021}
		M Gruneisen, M. Eickhoff, S. Newey, K. Stoltenberg, J. Morris, M. Bareian, and et al, Adaptive-optics-enabled quantum communication: A technique for daytime space-to-earth links, Phys. Rev. Appl. 16(1), 014067 (2021).
		\bibitem{Twiss1957}
		R. Twiss and R. Brown, The question of correlation between photons in coherent beams of light, Nature, 179(4570), 1128-1129 (1957).
		\bibitem{Zernike1938}
		F. Zernike, The concept of degree of coherence and its application to optical problems, Physica,  5(8), 785-795 (1938).
		\bibitem{Liu2021}
		L. Liu, L. Qu, C. Wu, and et al, Improved spatial resolution achieved by chromatic intensity interferometry, Phys. Rev. Lett. 127(10), 103601 (2021).
		%\bibitem{Pittman1995}
		%Pittman T B, Shih Y H, Strekalov D V, et al. Optical imaging by means of two-photon quantum entanglement[J]. Physical Review A, 1995, 52(5): R3429.
		\bibitem{Zhang2005}
		D. Zhang, Y. Zhai, L. Wu, and et al, Correlated two-photon imaging with true thermal light, Opt. lett. 30(18), 2354-2356 (2005).
		\bibitem{Valencia2006}
		A. Valencia, G. Scarcelli, M. D’Angelo, and et al, Two-photon imaging with thermal light, Phys. Rev. Lett. 94(6), 063601 (2005).
		\bibitem{Erkmen2010}
		B. Erkmen and J. H. Shapiro, Ghost imaging: from quantum to classical to computational, Adv. Opt. Photonics, 2(4), 405-450 (2010).
		\bibitem{Shapiro2012}
		J. H. Shapiro and R. W. Boyd, The physics of ghost imaging, Quantum Inf. Process, 11(4), 949-993 (2012).
		\bibitem{Yu2016}
		H. Yu, R. Lu, S. Han S, and et al, Fourier-transform ghost imaging with hard X rays, Phys. Rev. Lett. 117(11), 113901 (2016).
		\bibitem{Lis2018}
		S. Li, F. Cropp, K. Kabra, and et al, Electron ghost imaging, Phys. Rev. Lett. 121(11), 114801 (2018).
		\bibitem{Hodgman2019}
		S. Hodgman, W. Bu, S. Mann, and et al, Higher-order quantum ghost imaging with ultracold atoms, Phys. Rev. Lett. 122(23), 233601 (2019).
		\bibitem{Shapiro2008}  
		J. H. Shapiro, Computational ghost imaging, Phys. Rev. A 78(6), 061802 (2008).
		
		\bibitem{Hardy2013}
		N. D. Hardy and J. H. Shapiro, Computational ghost imaging versus imaging laser radar for three-dimensional imaging, Phys. Rev. A  87(2), 023820 (2013).
		
		\bibitem{Chen2009}
		X. Chen, Q. Liu, K. Luo, and et al, Lensless ghost imaging with true thermal light, Opt. lett. 34(5): 695-697 (2009).
		
		\bibitem{Cheng2009}  
		J. Cheng, Ghost imaging through turbulent atmosphere, Opt. Exp. 17(10), 7916-7921 (2009).
		\bibitem{Dixon2011}
		P. Dixon, G. Howland, K. W. Chan, and et al, Quantum ghost imaging through turbulence, Phys. Rev. A 83(5), 051803 (2011).
		\bibitem{Chan2011}
		K. W. Chan, D. Simon, A. Sergienko, and et al, Theoretical analysis of quantum ghost imaging through turbulence, Phys. Rev. A 84(4), 043807 (2011).
		\bibitem{Hardy2011}
		N. Hardy and J. H. Shapiro, Reflective ghost imaging through turbulence, Phys. Rev. A 84(6), 063824 (2011).
		\bibitem{Franson1992}	
		J. Franson, Nonlocal cancellation of dispersion, Phys. Rev. A, 45(5), 3126 (1992).
		\bibitem{Shapiro2010}
		J. H. Shapiro, Dispersion cancellation with phase-sensitive Gaussian-state light, Phys. Rev. A, 81(2), 023824 (2010).
		\bibitem{Sensarn2010}
		S. Sensarn, G. Yin, and S. Harris, Observation of nonlocal modulation with entangled photons, Phys. Rev. lett. 103(16), 163601 (2009).
		\bibitem{Torres2012}
		J. Torres and A. Friberg, Shaping the ultrafast temporal correlations of thermal-like photons, Phys. Rev. Lett, 109(24), 243905 (2012).
		\bibitem{Nodurft2020}
		I. Nodurft, S. Shringarpure, B. Kirby, T. Pittman, and J. Franson, Nonlocal dispersion cancellation for three or more photons. Phys. Rev. A, 102(1), 013713 (2020).
		\bibitem{Defienne2019}
		H. Defienne, M. Reichert, and J. Fleischer, Adaptive quantum optics with spatially entangled photon pairs, Phys. Rev. Lett. 121(23), 233601 (2018).
		\bibitem{Lib2020}
		O. Lib, G. Hasson, and Y. Bromberg, Real-time shaping of entangled photons by classical control and feedback, Sci. Adv. 6(37), eabb6298 (2020).
		\bibitem{Black2020}
		A. Black, E. Giese, B. Braverman, N. Zollo, S. Barnett, and R. W. Boyd, Quantum nonlocal aberration cancellation, Phys. Rev. Lett. 123(14), 143603 (2019).
		\bibitem{Bhusal2021}
		N. Bhusal, S. Lohani, C. You, M. Hong, J. Fabre, P. Zhao, and et al, Spatial mode correction of single photons using machine learning. Adv. Quantum Technol. 4(3), 2000103 (2021).
		\bibitem{Valencial2021}
		N. Valencia, S. Goel, W. McCutcheon, H. Defienne, and M. Malik, Unscrambling entanglement through a complex medium, Nat. Phys. 16(11), 1112-1116 (2020).
		\bibitem{Altmann2018}
		Y. Altmann, S. McLaughlin, M. Padgett, and et al, Quantum-inspired computational imaging. Science, 361(6403), eaat2298 (2018).
		\bibitem{Shapiro1974}
		J. H. Shapiro, Normal-mode approach to wave propagation in the turbulent atmosphere, Appl. Opt. 13(11), 2614-2619 (1974).
		\bibitem{Roddier}
		F. Roddier, Adaptive optics in astronomy[M], Cmbridge University Press, (2010) Chaper 4.
		\bibitem{Muller1974}
		R. Muller and A. Buffington. Real-time correction of atmospherically degraded telescope images through image sharpening, JOSA,  64(9), 1200-1210 (1974).
		
		\bibitem{Conkey2012}
		D. Conkey, A. Brown, A. Caravaca-Aguirre, and et al, Genetic algorithm optimization for focusing through turbid media in noisy environments, Opt. Exp. 20(5), 4840-4849 2012.
		\bibitem{Yeminy2021}
		T. Yeminy and O. Katz, Guidestar-free image-guided wavefront shaping, Sci. Adv. 7(21), eabf5364 (2021).
		\bibitem{Hirose2022}
		M. Hirose, N. Miyamura, and S. Sato, Deviation-based wavefront correction using the SPGD algorithm for high-resolution optical remote sensing, Appl. Opt. 61(23), 6722-6728 (2022).
		\bibitem{Sun2019}
		S. Sun S, W. Liu, J. Gu, and et al, Ghost imaging normalized by second-order coherence, Opt. Lett. 44(24), 5993-5996 (2019).
		\bibitem{Yang2020}
		D. Yang, G. Wu, J. Li, and et al, Image recovery of ghost imaging with sparse spatial frequencies, Opt. Lett. 45(19), 5356-5359 (2020).
		\bibitem{Lane1992}
		R. Lane, A. Glindemann, and J. Dainty, Simulation of a Kolmogorov phase screen. Waves in random media, 2(3), 209 (1992).
		\bibitem{Schmidt2010}
		J. D. Schmidt, Numerical simulation of optical wave propagation: With examples in MATLAB[C], SPIE, 2010.
		
		\bibitem{Goodman2007}
		Goodman J W. Speckle phenomena in optics: theory and applications[M]. Roberts and Company Publishers, 2007.
		
		\bibitem{Shapiro1981} 
		Shapiro J H, Capron B A, Harney R C. Imaging and target detection with a heterodyne-reception optical radar[J]. Applied optics, 1981, 20(19): 3292-3313.
		
		\bibitem{Erkmen2010}
		Erkmen B I, Shapiro J H. Ghost imaging: from quantum to classical to computational[J]. Advances in Optics and Photonics, 2010, 2(4): 405-450.
		
		\bibitem{fried1982}
		Fried D L. Anisoplanatism in adaptive optics[J]. JOSA, 1982, 72(1): 52-61.
		\bibitem{Tyson2022} 
		Tyson R K, Frazier B W. Principles of adaptive optics[M]. CRC press, 2022.
		
		\bibitem{Erkmen2008}   
		Erkmen B I, Shapiro J H. Unified theory of ghost imaging with Gaussian-state light[J]. Physical Review A, 2008, 77(4): 043809.
		
		
		
		
		
		
		
		
		%\bibitem{Gbur2001}
		%Gbur, G., and Wolf, E. (2002). Spreading of partially coherent beams in random media. JOSA a, 19(8), 1592-1598.
		%\bibitem{Salem2003}
		%Salem, M., Shirai, T., Dogariu, A., and Wolf, E. (2003). Long-distance propagation of partially coherent beams through atmospheric turbulence. Optics Communications, 216(4-6), 261-265.
		%\bibitem{Cao2005}
		%Cao D Z, Xiong J, Wang K. Geometrical optics in correlated imaging systems[J]. Physical Review A, 2005, 71(1): 013801.
		%\bibitem{Scarcelli20061}
		%Scarcelli G, Berardi V, Shih Y. Phase-conjugate mirror via two-photon thermal light imaging[J]. Applied physics letters, 2006, 88(6): 061106.
		
		
		%\bibitem{Lane1992}
		%Lane R G, Glindemann A, Dainty J C. Simulation of a Kolmogorov phase screen[J]. Waves in random media, 1992, 2(3): 209.
		
		%\bibliography{ref}
		
	\end{thebibliography}
\end{document}